\documentclass[preprint,11pt]{elsarticle}
\usepackage{graphicx}
\usepackage{amssymb}
\usepackage{hyperref}

\begin{document}
\begin{frontmatter}

\title{ The Effect of Pressure on the Negative
Thermal Expansion  of Solid Methane}

\author[firstaddress]{Yu.A. Freiman}

\author[firstaddress]{V.V. Vengerovsky}

\author[secondaddress]{ Alexander F. Goncharov}


\address[firstaddress]{ B.Verkin Institute for Low Temperature Physics and Engineering
of the National Academy of Sciences of Ukraine, 47 Nauky Avenue, Kharkiv 61103, Ukraine}
\address[secondaddress]{Geophysical Laboratory, Carnegie Institution of Washington, 5251 Broad Branch Road NW, Washington D.C. 20015, USA}

\begin{abstract}
The effect of pressure on
the thermal expansion of solid CH$_4$ is calculated for the low
temperature region where the contributions from phonons and
librons can be neglected and only the  rotational tunnelling modes
are essential. The effect of pressure is shown to increase the
magnitude of the peaks of the negative thermal expansion and
shifts the positions of the peaks to the low-temperature region,
which goes asymptotically to zero temperature  with increasing
pressure. The Gruneisen thermodynamical parameter
for the rotational tunnelling modes is calculated. It is large,
negative, and increases in magnitude with rising pressure.
\end{abstract}


\end{frontmatter}



\section{Introduction}
At low temperatures the rotational degrees of freedom in solid
methane are quantum, which makes solid methane (as well as solid
hydrogen), the most interesting simple molecular solids. In
addition, the methane molecule has the spin and in the condensed
state methane is a mixture of three spin modifications, $A, T$ and
$E$, the total spins of which is 2, 1 and 0, respectively.

Solid methane has a rich phase diagram with rather unusual
orientation structures. The high temperature phase I is
orientationally disordered with nearly freely rotating molecules.
The I-II transition is driven by the octupole-octupole
interaction. The I-II and II-III phase boundaries were first found
in the calorimetric measurements by Shubnikov, Trapeznikova and
Milyutin \cite{Shubnikov1939}. Both CH$_4$ and CD$_4$ undergo
orientational phase transitions to the partially ordered phase II
at 20.4 K and 27.0 K, respectively.

The structure of phase II has been predicted by James and Keenan
\cite{James1959} in 1959 and confirmed for CD$_4$ using coherent
neutron scattering by Press in 1972 \cite{Press1972}. In phase II
there are eight sublattices. On two of them the molecular field
vanishes; the molecules on these sites do not show orientational
order. On the six ordered sublattices the molecules are
orientationally ordered.

When temperature decreases, at 22 K CD$_4$ shows a further
transition to a slightly tetragonally distorted phase, whereas
CH$_4$ remains in phase II down to the lowest temperatures. The
low-temperature II-III phase boundary was determined by Nijman and
Trappeniers \cite{Nijman1977}. They showed that phase II of CH$_4$
is reentrant, that is, the phase boundary is curved at low
pressure and does not touch the $P=0$ line, so in contrast to
solid CD$_4$ phase III of solid CH$_4$ exists only under pressure
where the II-III phase boundary goes to the $P=0$ line at 22 K.
Below this temperature, only phase III exists.  Such difference in
the phase diagrams of light and heavy methanes results from the
quantum effect of molecular rotation, since the rotational
constant $B_{\rm rot} = \hbar^2/2I$ of CH$_4$ is twice as large as
that of CD$_4$ (5.3 and 2.6 cm$^{-1}$, respectively). Structure of
the tetragonal phase was proposed in the x-ray study by
Prokhvatilov and Isakina \cite{Prokhvatilov1980}.

Fabre et al. \cite{Fabre1979,Fabre1982} studied the
low-temperature Raman spectra of solid CH$_4$ up to 9 kbar. The
discontinues observed in the intramolecular and lattice
vibrational spectra were indicative the existence of five phases
in the pressure range below  9 kbar: phase II (0 - 0.5 kbar),
phase III  (0.5 - 1.9 kbar), phase IV (1.9 - 4.9 kbar), and phase
V (above 4.9 kbar).

The potential barriers arising from intermolecular interactions,
hinder free rotation of the methane molecules. Each molecule has
several equivalent minima and may tunnel from one position to
another resulting in a tunnelling splitting of the rotational
levels. Due to this tunnelling, the 16-fold degenerate ground
state in phase II is split into a five-fold degenerate level of
$A$ symmetry, a nine-fold $T$ level, and a two-fold degenerate $E$
level.  The lowest state corresponds to the $A$ modification, and
in the case of noninteracting molecules is characterized  by the
rotational quantum number $J=0$. The rotational wave function of
such a state is spherically symmetric, and the molecule in this
state has zero octupole moment. The splittings $\Delta_{AT}$ and
$\Delta_{TE}$, the energy gaps between the ground state and the
$T$ level and between the $T$ and $E$ levels, respectively have
been calculated in Refs \cite{Huller1975,Huller1979,Yamamoto1977}
and measured by the inelastic neutron scattering  technique in
Refs.
\cite{Glattli1972,Press1975,Baciocco1987,Medina1979,Buchman1982,Prager1981}.

The rotational tunnelling states in phases II and III of the
compressed solid CH$_4$ were studied by the inelastic neutron
scattering techniques at pressures up to 1.8 kbar
\cite{Eckert1981,Prager1982,Thiery1985} and were found that the
tunnelling energies strongly depend on pressure. By increasing the
pressure from 0.6 to 1.8 kbar the tunnelling lines shift by about
a factor of 1.5.

If the barrier height is raised the splittings become smaller. A
hydrostatic compression of the solid will increase the barrier
height thereby reducing the tunnelling probability and hence
decreasing the tunnelling splittings $\Delta_{AT}$, $\Delta_{TE}$
The volume change of the crystal with changing temperature (in the
temperature range where the the contribution of the usual phonon
mechanism is negligible) is determined by the competition between
two factors. The contribution to free energy due to populating of
the rotational tunnel states of the ordered sublattices  on rising
temperature favors contracting of the lattice. The height of the
potential barriers separating equivalent minima then increases,
the magnitude of the tunnelling splitting decreases and the
crystal free energy decreases. This effect is counterbalanced by
the loss in elastic energy increasing with increased contraction,
which stabilizes the crystal volume at each temperature. Thus,
rotational tunnelling gives rise to the negative thermal
expansion.

\begin{figure}
    \begin{center}
        \includegraphics[scale=0.7]{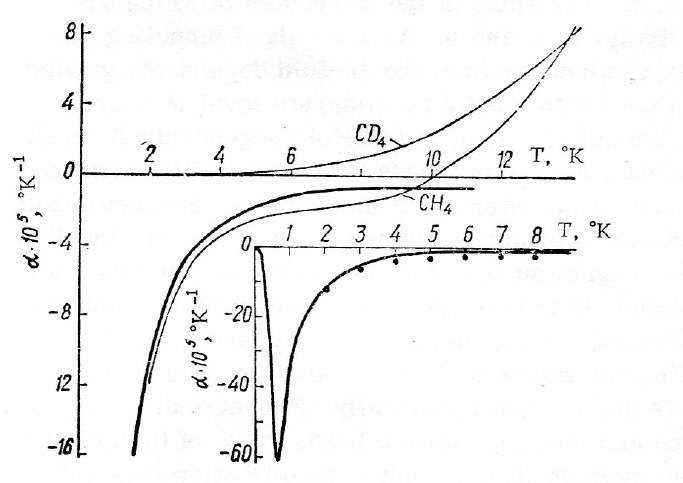}
    \end{center}
    \caption{Linear thermal expansion  coefficient of solid CH$_4$ and CD$_4$
    as a function of temperature (thin curves - experimental data \cite{Aleksandrovskii1978},
    solid curve  - theory \cite{Freiman1983}). Inset: the low temperature region
    (curve - theory \cite{Freiman1983}, points - experimental data  \cite{Aleksandrovskii1978}).
    (From Ref. \cite{Freiman1983}.)}
    \label{fig:Thermal expansion of solid methane at zero pressure}
\end{figure}

Heberlain and Adams \cite{Heberlain1970} found that the thermal
expansion coefficient (TEC) of solid methane becomes negative
below 8.7 K. Subsequent measurements
\cite{Aleksandrovskii1976,Aleksandrovskii1978} showed that as the
temperature is lowered the absolute value of the TEC being
negative continues to grow and at the lowest temperature achieved
in the dilatometric measurements (2 K) it is still far from a
maximum (Fig 1).

Yamamoto and Kataoka \cite{Yamamoto1970} proposed the following
qualitative explanation of this phenomenon. The lowest of the
splitted 16-fold ground state corresponds to the 5-fold state of
the $A$-modification, and in the case of noninteracting molecules
it is characterized by a rotational quantum number $J=0$. The
rotational wave function of such a state is spherically symmetric,
and a molecule in this state has zero octupole moment. In the
crystal the ground state is a superposition of states with
different values of $J$. However, the $A$-modification molecule
has a more spherically symmetrical spatial distribution and,
consequently, a lower value of the effective octupole moment than
$T$- and $E$-modification molecules. As the temperature is
reduced, the effective intermolecular octupole interaction
decreases and, as a result, to the crystal volume increases.
Conversion is essential for this mechanism to work, producing a
increase in the population of the $A$-modification ground state
with decreasing temperature. Aleksandrovskii et al.
\cite{Aleksandrovskii1976, Aleksandrovskii1978} demonstrated the
determining role of conversion in this mechanism.

Quantitative theory of the negative thermal expansion of solid
methane  at zero pressure based on the quantum rotational
tunnellimg was proposed in Ref. \cite{Freiman1983}. The effect of
pressure on the low-temperature thermal expansion is considered in
the present paper.

\section{Negative thermal expansion of compressed solid methane}

At sufficiently low temperatures when the phonon and libron
contributions can be neglected, crystal free energy can be written
in the form

$$
{\cal F}={\cal F}_{1}(V, T) + \frac{(V-V_0)^2}{2\chi V_0}.
\eqno(1)
$$
Here ${\cal F}_{1}$ is free energy associated with the tunnelling
states of the ordered sublattices:
$$
{\cal F}_{1}= -Nk_BT\ln[5 +
9\exp(-\Delta_{AT})/T)+2\exp(-\Delta_{AE})/T)]. \eqno(2)
$$
$N$ is the number of sites in the ordered sublattice ($N=
(3/4)N_0$), where $N_0$  is the number of molecules). The last
term in Eq. (1) is the lattice elastic energy, where $V(T)$ is the
volume, $T$ is temperature, $V_0$ is the crystal volume at $T=0$,
and $\chi$ is the isothermal compressibility.

We will not take the existence of the $E$ modification into
account in the further calculation. First, it has a lower
concentration and second, the contribution of this modifications
comes in the temperature region in which the phonon contribution
already appears.

So we will start from the free energy
$$
{\cal F} =  -Nk_BT\ln(5 + 9e^{-\Delta/T}) +
\frac{(V-V_0)^2}{2\chi V_0}, \eqno(3)
$$
where $\Delta =\Delta_{AT} $ is the gap between  the five-fold
degenerate ground state of the $A$ modification and nine-fold
levels of the $T$ modification.

The contribution of the tunnelling states into pressure is given
by the equation
$$ P= -\left(\frac{\partial{\cal F}}{\partial
V}\right)_T. \eqno(4)
$$

The volume dependence or ${\cal F}_1$ is connected with the
dependence of the energy  of the tunnelling states on the
internuclear distance. The equilibrium volume of the crystal at
temperature $T$ is determined by the condition for a minimum  in
free energy(considering the situation at zero pressure) $\partial
{\cal F}/\partial \Delta V  = 0$, $\Delta V = V(T) - V_0$. Since
the elastic energy is positive for any value of $\Delta V$, the
sign of the thermal expansion is determined  by the sign of the
derivative $\partial {\cal F}_1/\partial \Delta V$. It is easy to
see that $\partial {\cal F} /
\partial \Delta V > 0$  and the thermal expansion is negative.
Really, a reduction in crystal volume leads to an increase  in the
barrier separating equivalent minima, i.e. to a reduction in the
tunnelling probability  and in tunnelling splitting  and, as a
result of this, to a reduction in the magnitude of ${\cal F}_1$.
The loss in ${\cal F}_1$  is compensated by a gain in elastic
energy, which then determines the equilibrium volume $V(T)$ at
each temperatures. The considered mechanism leads to a negative
thermal expansion over the whole temperature range of existence of
phase II methane.  As the the temperature is raised , he
contribution of the usual  thermal expansion mechanism (phonon and
libron) starts to grow, and eventually the total thermal expansion
of the crystal becomes positive.

From Eq. (5) for the pressure we have the following equation:
$$
P = -9N\frac{e^{-\Delta/k_B\,T}}{5+ 9e^{-\Delta/k_B\,
T}}\frac{\partial \Delta}{\partial V} -\frac{V-V_0}{\chi
V_0}.\eqno(5)
$$

The coefficient of thermal expansion
$$
\beta_P = \frac{1}{V}\left(\frac{\partial V}{\partial T}\right)_P.
\eqno(6)
$$
Let us turn from the variables  $V, T$ to the variables $P, V$
using the Jacobian of the transformation $D(P, V)/D(T, V)$. As a
result we have
$$
\beta_P = -\left(\frac{\partial P}{\partial
T}\right)_V\,\frac{1}{V(\partial P/\partial V)}_T. \eqno(7)
$$

From Eq. (5) we have the following relations for the derivatives
$(\partial P/ \partial T)_V$ and $(\partial P)/\partial V)_T$:
$$
\left(\frac{\partial P}{\partial T}\right)_T =
-45N\left(\frac{1}{T}\right) \frac{\exp(-\Delta/k_B\,T)}
{[5+9\exp(-\Delta/k_B\,T)]^2}\frac{\partial \Delta}{\partial V}.
\eqno(8)
$$
$$
\left(\frac{\partial P}{\partial V} \right)_T = -\frac{1}{\chi
V_0} \left \{ 1 -45\frac{N\chi V_0}{\Delta}
\frac{(\Delta/k_B\,T)e^{-\Delta/k_B\,T}}{(5+9e^{-\Delta/k_B\,T})^2}\left(\frac{\partial
\Delta}{\partial V}\right)^2 + 9N\chi
V_0\frac{e^{-\Delta/k_B\,T}}{5+9e^{-\Delta/k_B\,T}}\frac{\partial^2\Delta}{\partial
V^2}\right\}^{-1}. \eqno(9)
$$
\begin{figure}
    \begin{center}
        \includegraphics[scale=1.2]{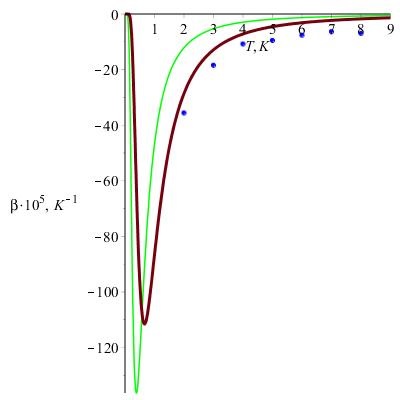}
    \end{center}
    \caption{The effect of pressure on the thermal expansion of solid methane.
   The volume expansion coefficient as a function of temperature: solid curve - zero pressure;
   thin curve - 850 bar, points - experimental data for $P= 0$  \cite{Aleksandrovskii1978}.}
    \label{fig: Pressure effect on thermal expansion}
    \end{figure}

Finally  for the coefficient of thermal expansion we have the
following relation:
$$
\beta_P = -45Nk_B\chi \left(\frac{V_0}{V} \right)
\frac{(\Delta/k_B\,T^2)e^{(-\Delta/k_B\,T)}}{[5+9e^{-\Delta/k_B\,T}]^2}\,\frac{\partial
\Delta}{\partial V}\,\times
$$
$$
\left \{ 1 -45\frac{N\chi V_0}{\Delta}
\frac{(\Delta/k_B\,T)e^{-\Delta/k_B\,T}}{(5+9e^{-\Delta/k_B\,T})^2}\left(\frac{\partial
\Delta}{\partial V}\right)^2 + 9N\chi
V_0\frac{e^{-\Delta/k_B\,T}}{5+9e^{-\Delta/k_B\,T}}\frac{\partial^2\Delta}{\partial
V^2}\right\}^{-1}. \eqno(10)
$$

Equation (10) should be supplemented by equations for $\Delta$ and
$\partial \Delta/\partial V$. We shall use for the further
calculations the dependence of the energy  of the tunnelling state
$\Delta$ at zero pressure on rotational barrier hight $U$,
obtained  by Huller and Raich \cite{Huller1979}:

$$
\Delta = \omega_0 e^{-\gamma U}, \eqno(11)
$$
were $U$ is the barrier hight in units of the rotational constant.
Taking into account that repulsive forces make the  largest
contribution to the  the derivative $\partial \Delta/ \partial V$
and assuming that there is a power law relation $U \sim r^{-n}$
and taking $n=15$ \cite{Nijman1977} we finally have
$$
(\partial \Delta /\partial V)_{P,T=0} = 5 (\Delta / V_0)(\gamma
U_0);\qquad (\partial^2\Delta/\partial V^2)_{P,T=0} = 25
(\Delta/V_0^2)(\gamma U_0)^2. \eqno(12)
$$
where $U_0$ is the reduced value of the barrier at zero pressure
and temperature. The effect of pressure on the thermal expansion of solid methane
can be seen from Fig. 2.

The sensitivity of the respective frequency spectrum to the
lattice expansion is described by the Gruneisen
parameter G
$$
G = \beta_P V/C_V\chi, \eqno(13)
$$
where $C_V$ describes the contribution of the respective modes to
the heat capacity $C_V= -T(\partial^2{\cal F}_{1}/\partial T^2)$. For the
rotational tunnelling modes from Eq (2) we have
$$
C_V^{\rm rot} =
45Nk_B\frac{(\Delta/k_B\,T)^2e^{-\Delta/k_B\,T}}{[5+9e^{-\Delta/k_B\,T}]^2}.
\eqno(14)
$$
$$
G(P)= \frac{V_0}{\Delta}\frac{\partial \Delta}{\partial V}
 \left \{ 1 -45\frac{N\chi V_0}{\Delta}
\frac{(\Delta/k_B\,T)e^{-\Delta/k_B\,T}}{(5+9e^{-\Delta/k_B\,T})^2}\left(\frac{\partial
\Delta}{\partial V}\right)^2 +
\frac{9N\,\chi\,V_0e^{-\Delta/k_B\,T}}{5+9e^{-\Delta/k_B\,T}}\frac{\partial^2\Delta}{\partial
V^2}\right\}^{-1}. \eqno(15)
$$

At zero pressure
$$
G(P=0) = - \frac{V_0}{\Delta}\left(\frac{\partial \Delta}{\partial
V}\right)_{P=0} = -5\gamma U_0 = -17.2.
\eqno(16)
$$
The pressure dependence of the thermodynamical Gruneisen parameter $G(P)$ is shown in Fig. 3.
\begin{figure}
    \begin{center}
        \includegraphics[scale=1.0]{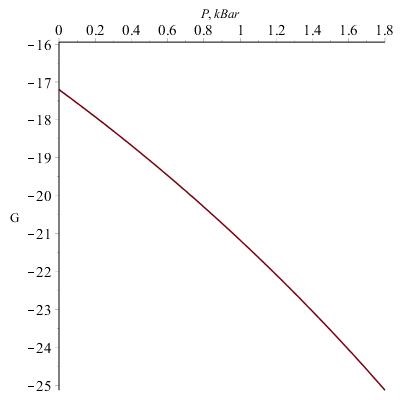}
    \end{center}

    \caption{Thermodynamical Gruneisen parameter of solid methane as a function of
    pressure.}
    \label{fig: Gruneisen parameter}
    \end{figure}




\section{Conclusions}

The effect of pressure on the thermal expansion of solid CH$_4$ is
calculated for the low temperature region where only the
rotational tunnelling modes are essential.It is shown that the
effect of pressure is quite unusual: the solid CH$_4$ becomes
increasingly quantum with rising pressure.

\end{document}